\documentstyle[11pt,aaspp4]{article}
\def\kms{km s$^{-1}$}
\def\solar{\ifmmode_{\mathord\odot}\else$_{\mathord\odot}$\fi}
\def\MM{$\cal{M}$\rm}
\def\MSUN{\MM\solar}

\def\etal{\it et al. \rm}
\def\Ha{H$\alpha$}

\def\kms{km-s$^{-1}$}
\newbox\grsign \setbox\grsign=\hbox{$>$} \newdimen\grdimen \grdimen=\ht\grsign
\newbox\simlessbox \newbox\simgreatbox
\setbox\simgreatbox=\hbox{\raise.5ex\hbox{$>$}\llap
     {\lower.5ex\hbox{$\sim$}}}\ht1=\grdimen\dp1=0pt
\setbox\simlessbox=\hbox{\raise.5ex\hbox{$<$}\llap
     {\lower.5ex\hbox{$\sim$}}}\ht2=\grdimen\dp2=0pt

\begin{document}
\centerline{}

\title{Keck Spectra of Pleiades Brown Dwarf Candidates and a Precise
Determination of the Lithium Depletion Edge in the 
Pleiades\footnote{Based on observations obtained at the W.M. Keck
Observatory, which is operated jointly by the University of California
and the California Institute of Technology} }
\author{John R. Stauffer}
\affil{Smithsonian Astrophysical Observatory, 60 Garden St., Cambridge, MA 02138}
\author{Greg Schultz} 
\affil{UCLA Astronomy and Astrophysics, Box 951562, Univ. of California,
Los Angeles, CA  90095-1562}
\author{J. Davy Kirkpatrick}
\affil{IPAC, MS 100-22, California Institute of Technology, Pasadena, CA 91125}

\vskip1.5truein
\noindent

\begin{abstract}

We have obtained intermediate resolution spectra of eleven candidate brown dwarf
members of the Pleiades open cluster using the Keck II telescope and LRIS
spectrograph.  Our primary goal was to
determine the location of the ``lithium depletion edge" in the Pleiades and
hence to derive  a precise age for the cluster.   
All but one of our 11 program objects have radial velocities appropriate for
Pleiades members, have moderately strong \Ha\ emission, and have spectral
types M6 to M8.5 as expected from their (R-I)$_c$\ colors.

We have constructed a color-magnitude diagram for the faint end of the Pleiades
main sequence, including only stars for which high S/N spectra in the region
of the lithium $\lambda$6708$\AA$\ absorption line have been obtained.  These
data allow us to accurately determine the Pleiades single-star lithium depletion
edge at I$_{c0}$\ = 17.80, (R-I)$_{c0}$\ = 2.20, spectral type = M6.5.  By reference to
theoretical evolutionary models, this converts fairly directly into an age
for the Pleiades of $\tau$\ = 125 Myr.  This is significantly older than 
the age that is normally quoted, but does agree with some other
recent estimates.  

\end{abstract}

\keywords{stars: low mass, brown dwarfs; 
open clusters and associations: individual (Pleiades) }

\section{Introduction}

The Pleiades was recognized in the 1980's as the best open cluster to
attempt to identify brown dwarfs (Stauffer \etal 1989; Jameson \& 
Skillen 1989) because of its fortuitous combination of proximity, youth
and richness.  A number of brown dwarf candidates were identified in those
and subsequent papers, usually on the basis of photometry obtained from
deep imaging surveys in two or more colors.   
A means to establish that a candidate brown dwarf was at least near, if not
necessarily below, the hydrogen burning mass limit 
was proposed by Rebolo, Mart\'{\i}n and
Magazz\`u (1992; hereafter RMM).  Below about 0.065 \MSUN, brown dwarfs should never
develop core temperatures sufficient for lithium ignition, and thus for
objects less massive than this we should find a lithium abundance the same
as for the interstellar medium from which the object formed, independent
of the object's age.  For slightly higher masses, lithium acts as an age scale
because the length of time it takes for the core to reach 2.5x10$^6$\ K is a
sensitive function of mass.  Because these stars are fully convective,
once the core temperature exceeds the necessary limit, the entire lithium
content of the star should be exhausted rapidly and thus be reflected in
an observable change in the photospheric lithium abundance.   The current
generation of theoretical models make specific predictions about the
time evolution of this lithium depletion boundary.  For example, D'Antona
\& Mazzitelli (1997) predict that at ages 30, 70 and 140 Myr, the lithium 
depletion edge should occur at 0.17 \MSUN, 0.09 \MSUN\ and 0.07 \MSUN,
respectively.   Other recent models by Baraffe \etal (1998)
and Burrows \etal (1997) make nearly identical predictions of the
variation of this lithium depletion boundary with age.   Indeed, Bildsten 
\etal (1997)
and others have argued that the age for an open cluster derived in this manner
should be better than
by any other method.

The initial attempts to detect lithium in 
very low mass open cluster members were not successful (RMM;
Marcy, Basri \& Graham 1994; hereafter MBG94).  However, Basri, Marcy \& Graham 
(1996; hereafter BMG96)  eventually detected lithium in 
PPL 15, a Pleiades brown dwarf candidate originally identified by
Stauffer, Hamilton \& Probst (1994).  Rebolo et al. (1996)
later showed that two other Pleiades members about 1 magnitude fainter than
PPL 15 also have strong lithium absorption.   BMG96 derived an age for the Pleiades
based on the presence of lithium in PPL 15 but the absence of lithium in
another only slightly brighter Pleiades member (HHJ3).  However, 
Basri \& Mart\'{\i}n (1998) have recently provided evidence that suggests
that PPL 15 is a nearly equal-mass spectroscopic binary and thus
its individual components would be $\sim$0.75 mag fainter.  This 
allows the location of the lithium boundary to be considerably
fainter than had been assumed.   The published data therefore
no longer
constrain the age of the Pleiades nearly as well as one would like because
too few stars have been measured spectroscopically in the  magnitude
range of interest.

In this paper, we report on the results of a program to determine
lithium abundances for a number of candidate Pleiades members with
apparent magnitudes chosen to bracket the possible magnitude range
within which the lithium depletion boundary might be located. 

\section{Observations and Data Reduction}

We obtained spectra of our Pleiades brown dwarf candidates with the Keck II
Low-Resolution Imaging Spectrograph (LRIS) on the nights of Dec. 3-5, 1997.
We used a 1200 l/mm grating and a one arcsecond slit.  The resultant
dispersion was 0.63 $\AA$/pixel, and the measured spectral resolution was
about 2.0 $\AA$.  The wavelength region observed was $\lambda\lambda$
6375 - 7650 $\AA$.  

The primary source from which we selected target objects was the recently
completed deep imaging survey of the Pleiades conducted by Bouvier 
\etal (1998).  Bouvier \etal
identified 18 objects from that survey as good candidate Pleiades
brown dwarfs based on their location in an I$_c$\ vs. (R-I)$_c$ color-magnitude
diagram.    We obtained Keck spectra of seven of those candidates, with
I magnitudes chosen to bracket the expected location of the lithium
depletion edge.  In addition, we obtained spectra of one target each
from four other Pleiades brown dwarf imaging surveys (Stauffer \etal
1989; Schultz 1998; Stauffer \etal 1998a; Zapatero-Osorio \etal 1997),
again selecting objects whose I magnitudes indicated they would help
define the lithium depletion edge.

To first order, all of our spectra look the same - they have strong TiO
and VO molecular bands indicative of spectral types M6 to M8, and all of
the stars we consider to be likely Pleiades members have moderately
strong \Ha\  emission.  They do show differences in their lithium
abundances, however.  Figure 1
shows a close-up of the lithium $\lambda$6708$\AA$\ region for 
three increasingly cooler Pleiades
members, illustrating that at M6.5 we do not detect lithium at good S/N,
that lithium is detected as a weak feature at M7 and that at later
spectral types it is generally seen quite strongly in absorption.

We have used our spectra to provide the following quantitative
information for our target objects:  (a) \Ha\ and lithium 
equivalent widths; (b) spectral indices
which measure the continuum shape from 7000$\AA$\ to 7500$\AA$\ (PC2 - see Mart\'{\i}n 
\etal 1996) and the depth of the VO band centered around 7450$\AA$\
(Kirkpatrick et al. 1995); and (c) radial velocities derived both from
gaussian fits to the \Ha\ emission feature (the strongest, best defined
feature in the spectra) and from cross-correlation of the region from
6580$\AA$\ to 7050$\AA$\ versus GL905 - a dM5.5 standard.

Table 1 provides the relevant photometric and spectroscopic data for
our target stars.  Columns 2 and 3 provide I$_{c0}$\ and (R-I)$_{c0}$\ 
photometry from Bouvier
\etal\ (1998) for the CFHT stars and from the papers cited above for the
other stars, where the ``0" subscript indicates that we have corrected
the observed photometry 
for an assumed reddening of A$_I$ = 0.06 and
E(R-I) = 0.03 (Stauffer \& Hartmann 1987).  
Columns 4 and 5 are the PC2 and VO spectral indices, while
column 6 is our spectral type derived from these indices using the
stars in Kirkpatrick et al. (1995) and Mart\'{\i}n \etal (1996) as
standards.  We have used Gliese catalog M dwarfs to define a linear
relation between PC2 and (R-I)$_c$:  (R-I)$_c$\ = 0.746 x PC2 + 0.65 (see
Stauffer et al. 1998b for details).  These spectroscopic R-I colors are
provided in column 7 of Table 1.  The last two columns of the table are
the measured lithium and \Ha\ equivalent widths.
We will report the radial velocities and some other information derived
from these spectra, including lithium abundances, in a subsequent paper
(Stauffer \etal 1998b).

In the next section, we will present a color-magnitude diagram for the
late-type Pleiades members for which there exists high quality lithium
equivalent width estimates.  We cannot do this using measured
photometric colors because the appropriate photometry does not exist
[that is, some of the stars have (R-I)$_c$\ colors, 
some have I-K colors but there is no single color where all of the
stars have measured photometry].  However, all of the stars have (R-I)$_c$\
colors estimated from spectroscopy or have spectral indices from which
(R-I)$_c$\ colors can be inferred.  We therefore will use
spectroscopic (R-I)$_c$\ colors in that figure.  Comparison of the photometric
and spectroscopic estimates of (R-I)$_c$\ for the seven CFHT stars in
Table 1 suggests that this is an acceptable procedure.

It is also necessary to know whether the objects we have observed are
Pleiades members or not.  As discussed in Bouvier \etal (1998),
the probability that a photometric candidate from the CFHT survey
will indeed be a Pleiades member, as estimated simply from the
expected field star contamination given the area and depth surveyed, is
fairly good - $\sim$70\%\ or better.  If in addition the object has a spectral
type that agrees with the photometric color, has \Ha\ emission strength
that matches what is expected for very low mass Pleiades members
(Stauffer, Liebert and Giampapa 1995), and has detected lithium - 
indicating an age $<$ 1 Gyr for stars in this spectral type range - 
then we believe that the object is almost certainly a Pleiades
member.  Non-detection of lithium fainter than the I$_c$\ magnitude where
other Pleiades members have detected lithium would suggest the object
is a non-member.  For brighter candidate Pleiades members, lithium is
not useful but radial velocities can provide further evidence in
support of
Pleiades membership.  In fact, all of the stars in Table 1 except
CFHT-PL-14 have radial velocities compatible with Pleiades membership
within the expected accuracy of our measurements ($\sim$4 \kms, 1$\sigma$,
for the velocities derived from \Ha, $\sim$8 \kms, 1$\sigma$, for the
cross-correlation velocities). 
CFHT-PL-14 also lacks \Ha\ emission and has no measurable lithium
feature.  We therefore assume that all of the stars in Table 1 are
Pleiades members except for CFHT-PL-14, which we assume is a
non-member.

\section{Discussion}

The goal of this project was to precisely determine the location of the
lithium depletion edge in the Pleiades, and hence to determine an
accurate age for the cluster.  Figure 2 shows a color-magnitude diagram
for very low mass Pleiades members for which lithium data are
available.  The lithium data are from Oppenheimer \etal (1997 =
Opp97), MBG94, BMG96, Rebolo \etal (1996 = Reb96) and from this paper. 
The (R-I)$_c$\ colors for the Opp97 stars are from Stauffer \etal (1995),
where we have converted these spectroscopic (V-I)$_c$\ colors 
to (R-I)$_c$\ via
a relation derived from the Gliese catalog M dwarfs in Leggett (1992). 
The (R-I)$_c$\ colors for some stars in Reb96 and BMG96 are from PC2
indices provided by Mart\'{\i}n \etal (1996) and our calibration of PC2
vs. (R-I)$_c$.  Two points for CFHT-PL-15 are shown: (R-I)$_c$\ = 2.24 as
derived from its PC2 index, and (R-I)$_c$ = 2.41 as derived from its
VO index.  Using either color, 
CFHT-PL-15 lies below the main sequence defined by the other stars,
possibly indicating that it is a non-member.  Because it has detected
lithium and a radial velocity compatible with Pleiades membership, we
prefer to believe it is a member and that the inferred color is
anomalous (either intrinsically or due to measurement error).  Our
primary conclusions are unaffected by how we interpret this star. 
Finally, the dashed line in Figure 2 is a field star ZAMS derived from
the M dwarf photometry provided by Leggett (1992).

The location of the stars in Figure 2 show a well-defined correlation
with the measured lithium equivalent widths.  Bluer than (R-I)$_c$\ = 2.2,
none of the stars have detected lithium while redder than that color all of
the measured stars have lithium.  Similarly, fainter than I$_{c0}$\ = 17.8
all the stars have detected lithium, whereas brighter than that limit
all but one of the stars have no lithium.  Finally, there is a trend
for lithium equivalent width to increase going to lower inferred mass
(i.e. fainter and redder).  We further believe that the dispersion in I
magnitude at a given color is to a large extent real and is primarily
an indication that some of the observed stars are photometric binaries.
In particular, we suggest that HHJ6 (I$_{c0}$\ = 16.93, R-I$_c$\ = 2.18), 
Lick-PL1, PPL1, and CFHT-PL-12 are good
candidates to be photometric binaries.  CFHT-PL-12 also has considerably
stronger \Ha\ emission than the other Pleiades stars observed - possibly
indicating it is a short-period binary or that we have observed it during
a flare.   The attribution of PPL1 as a nearly equal mass binary would
explain why it has
strong lithium absorption despite an I$_c$\ magnitude equal to or brighter
than two other cluster members with no detected lithium.  An
alternative explanation would be that there is a significant age
spread in the Pleiades and these four over-luminous stars are the
youngest in our sample.  By comparison to theoretical models
(e.g. D'Antona \& Mazzitelli 1997), their displacement about 0.5 mag
above other stars of the same color would require them to be more
than 50 Myr younger than the other stars, which we believe is
unlikely (see, for example, discussions of this issue by Soderblom \etal
1993 and Steele \& Jameson 1995).

Based on the above interpretation of Figure 2, we determine that the
single star lithium depletion edge in the Pleiades is at I$_{c0}$\ = 17.8 
$\pm$ 0.1 or (R-I)$_{c0}$\  = 2.20 $\pm$ 0.05.  The uncertainty estimates
are not rigorous, and arise mostly from the uncertainties in the absolute
calibration of the photometry in Bouvier \etal (1998).  For theoretical evolutionary
models which incorporate realistic model atmospheres as their outer
boundary condition, and hence which can predict observational colors
and magnitudes for brown dwarfs, it is possible to convert directly the
empirical lithium depletion edge to an age estimate for the Pleaides.
In Figure 3, we plot the absolute I$_c$\ magnitude of the lithium depletion boundary
(defined here as the point where lithium has been depleted by a factor
of 100) for the most recent models of Baraffe \etal (1998) as a function
of age.  To place our empirically measured point into this diagram,
we assume that the cluster has (m-M)$_o$\ = 5.60 (r $\sim$ 130 pc), and
A$_I$\ = 0.06 (c.f. Pinsonneault \etal 1998), leading to M(I$_c$) = 12.2
$\pm$\ 0.15 for the lithium depletion boundary, where we have assumed
plausible but again unrigorous 1$\sigma$\ uncertainties of 0.1 mag for the
distance modulus and 0.03 mag for the extinction and that the uncertainties
add in quadrature.  The age derived in this
way is then 125 $\pm$\ 8 Myr, where the uncertainty estimate only comes
from propagating the 0.15 mag uncertainty of the boundary
through the model shown in Figure 3.

In order to try to assess the model dependence of this estimate, we have also 
made a similar calculation for theoretical evolutionary
models by Burrows \etal (1997) and D'Antona \& Mazzitelli (1997).   Those
models do not provide R and I magnitudes, so instead we have used the
(R-I)$_c$ color to estimate an I$_c$-band bolometric correction (we adopted
the Monet \etal (1992) BC$_I$ vs. (V-I)$_c$\ relation, and a conversion from 
(V-I)$_c$\ to (R-I)$_c$\ based on data in Leggett 1992). In that manner, we estimate that the
lithium depletion boundary in the Pleiades is at M(Bol) = 11.99.  Comparing
this number to the predictions of the
two theoretical models, we get an age for the Pleiades
of 130 Myr based on the D'Antona \& Mazzitelli (1997) calculations and
125 Myr for the Burrows \etal (1997) models. 

The most commonly quoted age for the Pleiades is of order 70-80 Myr
(Patenaude 1978; Mermilliod 1981).  However, models with a relatively large
amount of convective core overshoot can yield much larger ages, as was
originally shown by the models of the Padova group (c.f. Mazzei \&
Pigatto 1989, who derived an age for the Pleiades of 150 Myr).  Other
recent models give ages intermediate between these values (e.g. 100 Myr for
Meynet, Mermilliod \& Maeder 1993 and $\geq$\ 120 Myr for Ventura \etal
1998).  Given the disagreement over the age derived from the upper main
sequence turn-off, it is particularly useful to have an independent means to
derive the age from low mass stars.  Using
the lithium detection in PPL15 and the non-detection of lithium in HHJ3,
BMG96 estimated the age of the Pleiades to be 115 Myr.  Our new result pushes
this age even slightly older, but more importantly does so using many
more stars and thus provides a much better defined age from the lithium
data.

Finally, we note that we have chosen to use the ``traditional" distance
scale to the Pleiades (r $\sim$\ 130 pc).   That distance conflicts with
the new Hipparcos distance to the Pleiades of about 116 pc (van Leeuwen
\& Ruiz 1997; Mermilliod \etal 1997).  We have done this because we believe
that the Hipparcos distance for the Pleiades is not correct, as has been
discussed in Pinsonneault \etal (1998) and Soderblom \etal (1998).  
The zeroth order effect of simply using the Hipparcos distance
and keeping everything else the same would lead to an even older Pleiades
age of about 140 Myr.

\section{Summary and Implications} 

Using the Keck II telescope, we have obtained good intermediate-resolution
spectra of ten very low mass, probable members of the Pleaides cluster.  We
detect lithium in seven of these stars, bringing the total number of confirmed
lithium brown dwarfs in the Pleiades to ten.  By also having spectra of several
slightly brighter Pleiades members which do not show lithium,
we have accurately determined the lithium depletion edge in the Pleiades to be
at M(I$_c$)\ = 12.2.  Using three different theoretical evolutionary models, we
derive a ``lithium age" for the Pleiades of 125-130 Myr.  The age of the Pleiades
as derived from the upper main sequence turnoff is variously quoted as between
$\sim$75 Myr and $\sim$150 Myr.  If our new age as derived from location of
the lithium depletion boundary is correct, and if the Pleiades is normal in
some sense, our result suggests that the ages for many young open clusters
should be shifted towards the old end of the scale, in approximate accord
with the age-scale derived from the models of the Padova group (Mazzei \&
Pigatto 1988) or that of Ventura \etal (1998).

Because the Pleiades stars have been extensively studied, they have
served as one of the cornerstones in defining how rotational velocities,
lithium abundances, coronal activity, etc. evolve with time. For most of
those studies, an age of 75-100 Myr has been adopted for the Pleiades.  Shifting the
age of the Pleiades to 125 Myr (with or without shifting the ages of the other
well-studied nearby clusters) will require modifications to our view of
how these various properties evolve with time.  This in turn will then affect
the theoretical models that have been developed to explain these observed
trends.

\acknowledgments 
JRS acknowledges support from NASA Grants NAGW-2698 and NAGW-3690.
GS acknowledges support from NASA and NSF grants to UCLA.

\vfill
\eject
\begin{figure}
\figurenum{1}
\plotfiddle{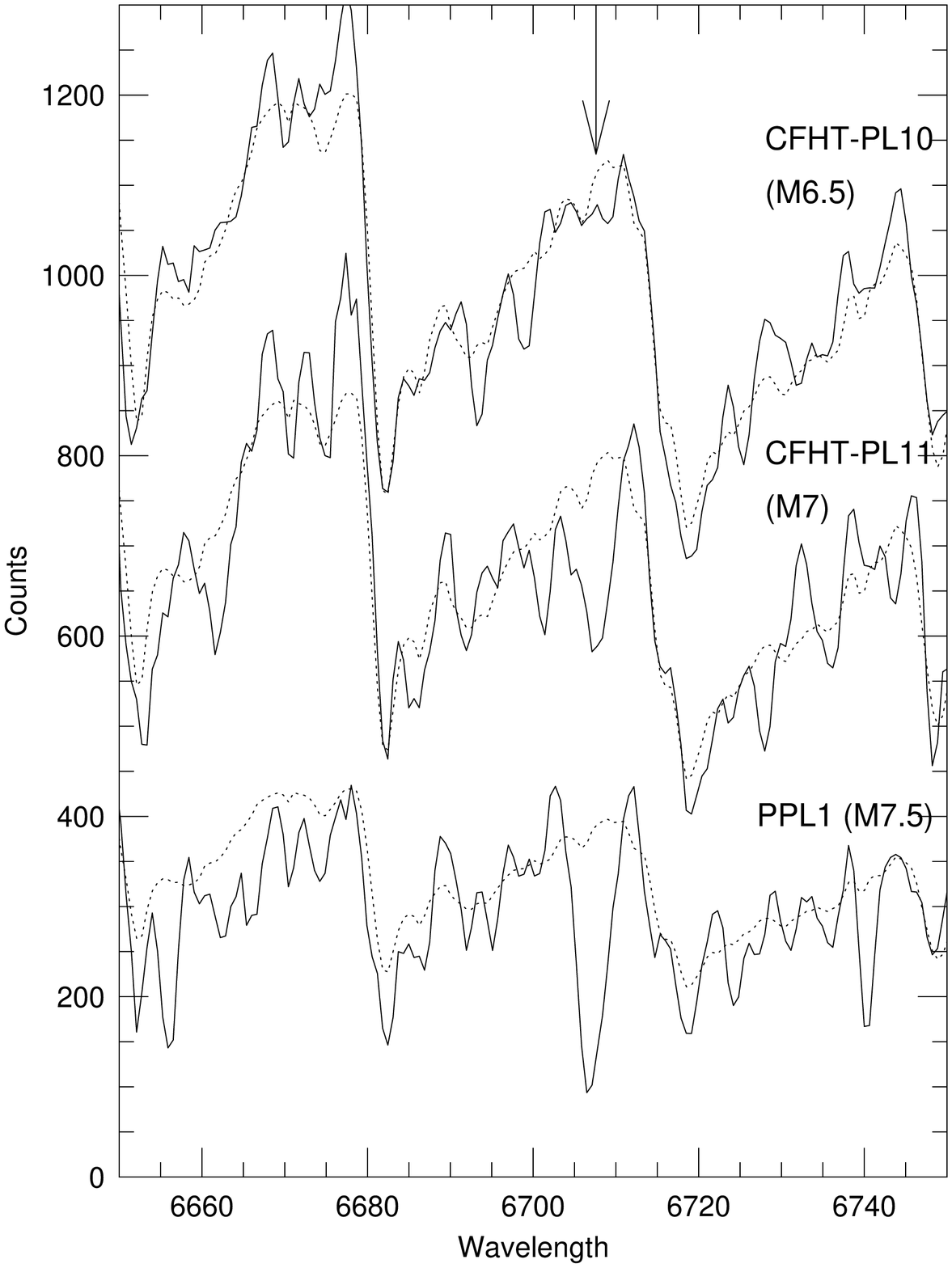}{4.5in}{0}{80}{65}{-250}{-50}
\caption{
Sample spectra of Pleiades brown dwarf candidates obtained
with the Keck II LRIS spectrograph.  The displayed wavelength region is
only a small portion of the full spectrum, selected in order to
highlight the lithium $\lambda$6708$\AA$\  region. The y-axis is correct for
CFHT-PL-10, while the spectra of the other two stars are offset relative to
CFHT-PL-10 to avoid having the spectra overlap.  The
dashed line is a spectrum of GL65AB, a field M6-M6.5 binary, assumed 
to have entirely depleted its initial lithium.
}
\end{figure}

\begin{figure}
\figurenum{2}
\plotfiddle{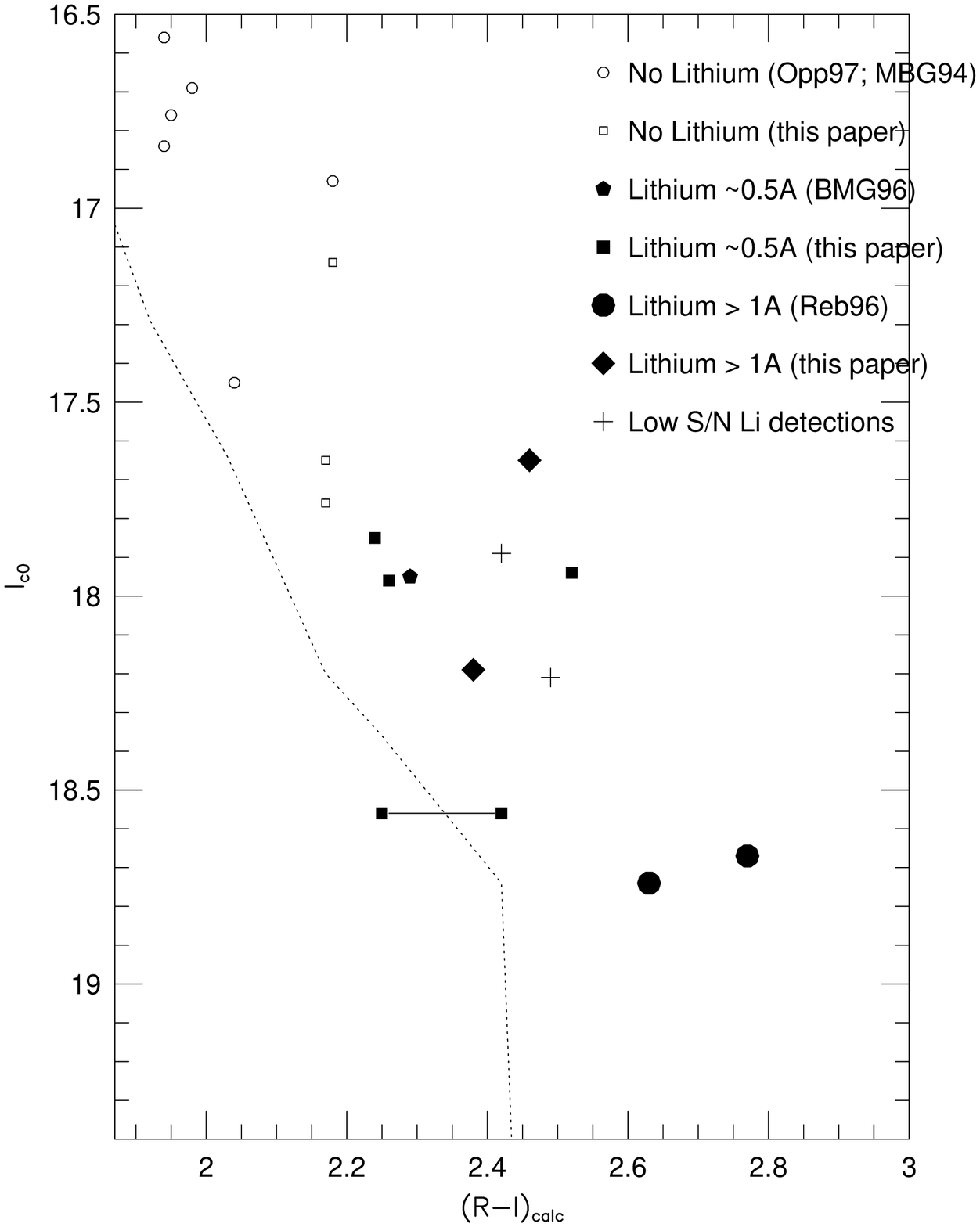}{4.5in}{0}{80}{65}{-250}{-50}
\caption{
Color-magnitude diagram showing only those very low mass
Pleiades members for which a good assessment of their lithium abundance 
can be made.  The dashed line is an empirical ZAMS shifted to Pleiades
distance.  The x-axis colors are inferred from
our spectroscopy (and so are reddening free); the y-axis has been corrected
for an assumed reddening of A$_I$\ = 0.06.  The two ``low-S/N detections"
are MHObd3 (in Table 1) and MHObd1 (see Stauffer \etal 1998a).
}
\end{figure}

\begin{figure}
\figurenum{3}
\plotfiddle{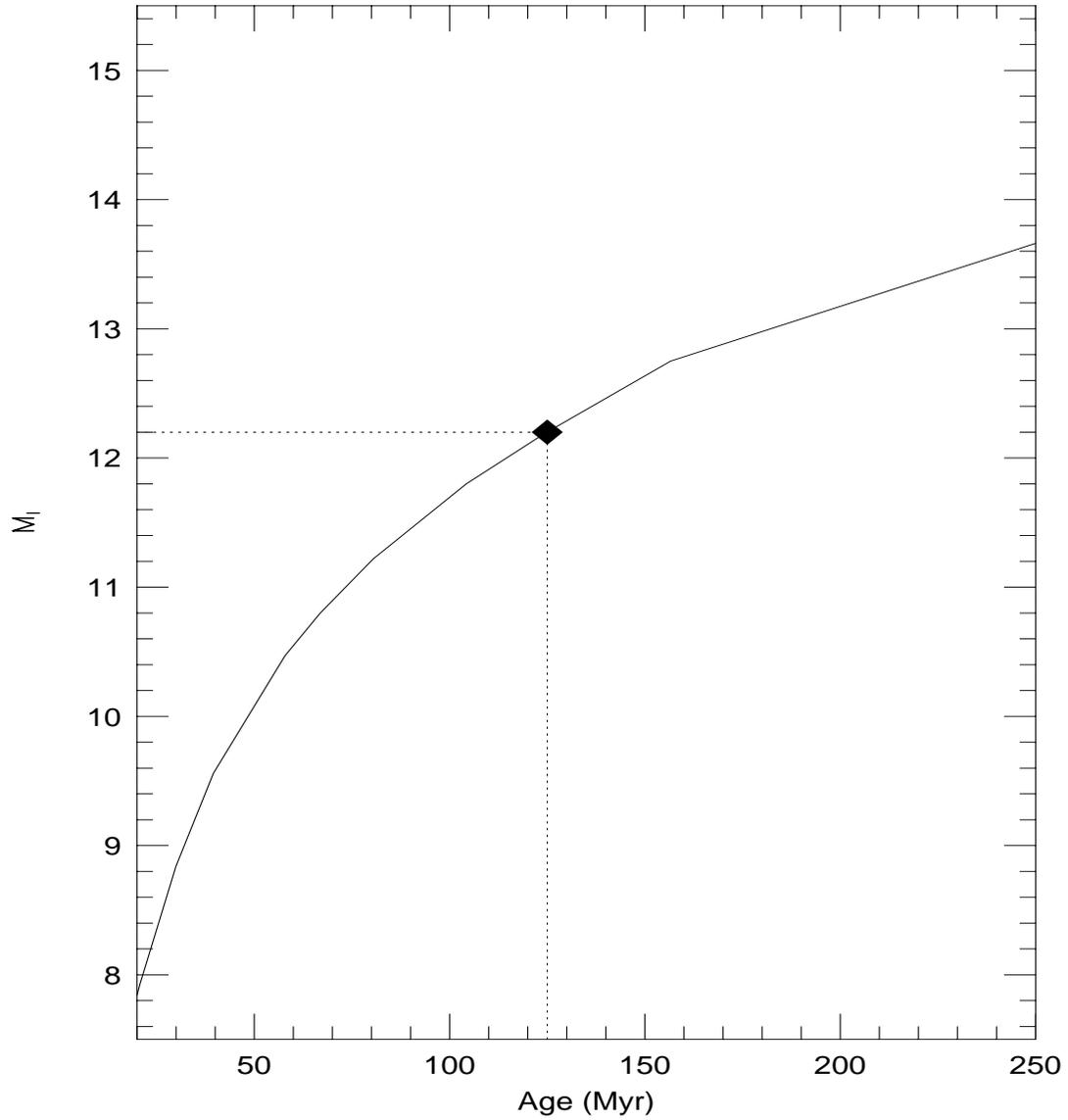}{4.5in}{0}{80}{65}{-250}{-50}
\caption{
Predicted absolute I$_c$\ magnitude where stars should have depleted
99\% of their initial lithium as a function of age based on the models
of Baraffe \etal (1998).   The diamond symbol marks our empirical location of this 
lithium depletion boundary in the Pleiades - indicating an age for the
cluster of 125 Myr.
}
\end{figure}

\begin{table}
\caption{Data for Pleiades Brown Dwarf Candidates}
\begin{tabular}{lcccccccc}
\tableline
      &           &               &     &    &         &                & EqW(Li) & EqW(H$\alpha$) \\
 Name &  I$_{c0}$ & (R-I)$_{c0}$  & PC2 & VO & Sp.Type & (R-I)$_{calc}$ & (\AA) & (\AA)  \\
\tableline
CFHT-PL-9  & 17.65 & 2.15 & 2.03 & 1.05 & M6.5 & 2.17 & $<$0.05 & 3.8 \\
CFHT-PL-10 & 17.76 & 2.18 & 2.03 & 1.04 & M6.5 & 2.17 & $<$0.05 & 5.4 \\
CFHT-PL-11 & 17.85 & 2.18 & 2.13 & 1.06 & M7   & 2.24 & 0.5     & 5.1 \\
CFHT-PL-12 & 17.94 & 2.44 & 2.51 & 1.11 & M8   & 2.52 & 0.8     & 17.2 \\
CFHT-PL-13 & 17.96 & 2.18 & 2.16 & 1.06 & M7   & 2.26 & 0.6     & 8.7 \\
CFHT-PL-14 & 17.98 & 2.14 & 1.87 & 1.03 & M6   & 2.05 & $<$0.1  & 0. \\
CFHT-PL-15 & 18.56 & 2.31 & 2.14 & 1.07 & M7   & 2.25 & 0.5     & 7.0 \\
Roq 13     & 18.19 & .... & 2.31 & 1.08 & M7.5 & 2.38 & 2.5     & 4.8 \\
MHObd3     & 18.21 & .... & 2.46 & 1.11 & M8   & 2.49 & 0.8::   & 5.9 \\
PPL 1      & 17.65 & .... & 2.42 & 1.11 & M7.5 & 2.46 & 2.4     & 6.5 \\
Lick PL1   & 17.14 &  .... & 2.05 & 1.03 & M6.5 & 2.18 & $<$0.1  & 5.3 \\
\tableline 
\end{tabular}
\end{table}

\end{document}